\documentclass[AMA,STIX1COL,doublespace]{WileyNJD-v2}
\usepackage{moreverb}
\usepackage{setspace}

\usepackage{hyperref}
\hypersetup{
	colorlinks   = true, 
	urlcolor     = blue, 
	linkcolor    = blue, 
	citecolor   = blue 
}

\newcommand\BibTeX{{\rmfamily B\kern-.05em \textsc{i\kern-.025em b}\kern-.08em
T\kern-.1667em\lower.7ex\hbox{E}\kern-.125emX}}

\articletype{Main Paper}%

\received{<day> <Month>, <year>}
\revised{<day> <Month>, <year>}
\accepted{<day> <Month>, <year>}

\begin{document}

\title{An Empirical Bayes Robust Meta-Analytical-Predictive Prior to Adaptively Leverage External Data}

\author[1]{Hongtao Zhang*}

\author[2]{Yueqi Shen}

\author[1]{Alan Y Chiang}

\author[3]{Judy Li}

\authormark{Zhang \textsc{et al}}

\address[1]{\orgdiv{Global Biometrics and Data Sciences}, \orgname{Bristol Myers Squibb}, \orgaddress{\state{Berkeley Heights, New Jersey}, \country{USA}}}

\address[2]{\orgdiv{Department of Biostatistics}, \orgname{University of North Carolina}, \orgaddress{\state{Chapel Hill, North Carolina}, \country{USA}}}

\address[3]{\orgdiv{Global Biometrics and Data Sciences}, \orgname{Bristol Myers Squibb}, \orgaddress{\state{San Diego, California}, \country{USA}}}

\corres{*Hongtao Zhang. \email{squallteo@gmail.com}}


\abstract[Abstract]{
We propose a novel empirical Bayes robust MAP (EB-rMAP) prior to adaptively leverage external/historical data. Built on Box's prior predictive p-value, the EB-rMAP prior framework balances between model parsimony and flexibility through a tuning parameter. The proposed framework can be applied to binomial, normal, and time-to-event endpoints. Computational aspects of the framework are efficient. Simulations results with different endpoints demonstrate that the EB-rMAP prior is robust in the presence of prior-data conflict while preserving statistical power. The proposed EB-rMAP prior is then applied to a clinical dataset that comprises of ten oncology clinical trials, including the perspective study. 
}



\maketitle


\section{Introduction}
External/Historical control information plays an increasingly important role at the design and analysis stage of a new clinical trial. For clinical trial sponsors, such information can lead to a more efficient study in terms of shorter duration and reduced cost. From the ethical perspective, it is possible that the number of patients assigned to the control arm can be lowered without compromising the overall statistical power. Bayesian methods are naturally suited for leveraging external control by utilizing the prior information derived therefrom. There are two popular classes of methods proposed: power prior approaches \cite{IbrahimEtal_2015} and meta-analytical-predictive (MAP) prior approaches \cite{NeuenschwanderEtal_2010}. The power prior approach assumes an identical parameter (e.g., mean response or rate) among external and current data, and discount the former with a discounting factor $a_0$. In contrast, such parameters arise from a common distribution in MAP prior via the exchangeability assumption, in which between-trial heterogeneity is controlled by (hyper-)parameters. A comparative review of statistical methods has been conducted by Viele and colleagues \cite{VieleEtal_2014}. 

It is well-acknowledged that utilizing external information possesses the risk of erroneous conclusion and bias when historical and current data are heterogeneous, a phenomenon also known as prior-data conflict. This caveat has arguably been the most important cause of reluctance to consider incorporating external information in trials. Therefore, a critical consideration is the robustness of the statistical models in the presence of prior-data conflict. A popular two-step procedure is test-then-pool. This approach first assesses the congruence between historical and current data via a hypothesis test of equality, and only pools the data if the null hypothesis is not rejected. On the other hand, the original MAP prior approach \cite{NeuenschwanderEtal_2010} belongs to a class of dynamic borrowing approaches \cite{VieleEtal_2014}. As the name suggests, such approaches determine the extent of borrowing dynamically based on the congruence between historical and current data. Further, the robust MAP (rMAP) prior \cite{SchmidliEtal_2014} was proposed to be a more robust extension of the MAP prior. The rMAP prior introduces a vague prior component and a mixture weight $w_V$ into the MAP prior framework. In particular, $w_V$ is pre-specified based on the anticipated likelihood of prior-data conflict. The rMAP prior approach regulates the amount of borrowing through the mixture weight $w_V$. When $w_V$ is closer to 1, rMAP prior is more dominated by the robust prior which is less informative, but more robust in the presence of increasing prior-data conflict; or vice versa when $w_V$ is closer to 0. 

Another desirable feature of a robust borrowing method is the ability to adjust for the amount of borrowing in a data-dependent and objective manner. In other words, the extent of borrowing is determined by the congruence between historical and current data, as opposed to a pre-specified parameter value such as $w_V$ in rMAP prior. This is because it is generally difficult to predict the likelihood of prior-data conflict at the trial design stage when no current data is available. With such methodology, the amount of uncertainty can be reduced in protocol development, which may facilitate the acceptance of external control by both sponsors and regulators. Recent developments in this direction include some extensions of the power prior paradigm \cite{GravestockHeld_2017, NikolakopoulosEtal_2018, BennettEtal_2021}, a Bayesian semiparametric MAP prior \cite{HupfEtal_2021_MAP_DP} and an empirical Bayes MAP prior \cite{LiEtal_2016}. In particular, the Bayesian semiparametric MAP prior approach uses the Dirichlet process prior to adaptively learn the relationship between historical and current data. The empirical Bayes MAP prior discounts or amplifies the impact of historical data based on a parameter determined by the the congruence between historical and current data. 
{\color{black}
Some early empirical Bayes applications to further frequentist goals while “double dipping the data” (first to determine the prior, and then in computing the posterior) have been critically reviewed by several authors. See, for example, Lindley (1983) \cite{Lindley_1983}. Nevertheless, the value of borrowing strength from similar but independent experiments have been appreciated. The review article by Carlin and Louis \cite{CarlinLouis_2000} and the cited references offer an excellent summary of applications for empirical Bayes methods in the biomedical sciences. 
}
Most aforementioned data-dependent borrowing methods focused on a particular type of endpoint such as binary or normal. While their respective frameworks may be applicable to other types of endpoints, the implementations of such extensions may not always be trivial. 

In this research, we propose a novel empirical Bayes robust MAP (EB-rMAP) prior to adaptively leverage external/historical data. As an MAP-prior-based method, EB-rMAP prior is ideal for handling multiple historical data sources, which could be challenging in empirical Bayes power prior methods \cite{NikolakopoulosEtal_2018}. Built on the Box's prior predictive p-value \cite{Box_1980}, the EB-rMAP prior framework balances between model parsimony and flexibility by introducing only one additional tuning parameter. The computation can be conducted through existing software packages and therefore is highly efficient. The unified framework can be seamlessly applied to most popular types of endpoint, including binary, normal and time-to-event (TTE). 

In Section 2, we firstly briefly review the MAP and rMAP priors and then introduce the EB-rMAP prior framework, with special considerations for the TTE outcome. We conduct simulation studies for binary, normal and TTE endpoints in Section 3. The TTE data from Roychoudhury and Neuenschwander \cite{RoychoudhuryNeuenschwander_2020} are re-analyzed with EB-rMAP prior in Section 4. Some future research topics of interest are discussed in the Concluding Remarks section. 

\section{Method}
\subsection{Original, Robust and Empirical Bayes MAP Priors}
We use the binary endpoint for illustration purposes. Denote $Y_C$ and $\theta_C$ the data and parameter of interest for the current control arm respectively. In this case, $\theta_C$ is the log-odds of response rate $p_C$ and $Y_C$ is the number of responders. Let $d_0 = \{Y_1,\cdots,Y_H\}$ be the control data from $H$ historical studies, arising from respective binomial distributions with probabilities $p_1,\cdots,p_H$. A hierarchical model is formed in original MAP prior \cite{NeuenschwanderEtal_2010}. 
\begin{enumerate}
	\item (Sampling models of data) For $i = 1,...,H, Y_i|p_i \sim Bin(n_i, p_i)$, where $n$ is the sample size. The sampling models are trial-specific. 
	\item (Exchangeability) A common distribution is assumed for log-odds: $\theta_1,..., \theta_H | \mu_C, \sigma_C \sim N(\mu_C, \sigma_C^2)$.
	\item (Hyper-priors) The mean parameter $\mu_C$ usually follows a non-informative prior. On the other hand, the exchangeability parameter $\sigma_C^2$ measures the between-trial heterogeneity and thus regulates the extent of borrowing. Common choices of priors are inverse-gamma distribution for $\sigma_C^2$ and half-normal/t families for $\sigma_C$.  
\end{enumerate}

The original MAP prior, $f_{\scalebox{0.5}{MAP}}(\theta_C|d_0)$, derived from the hierarchical model is conditional on historical data only. It can be used at the design stage of the current trial, when current data is unobserved. Further denote $D_1$ and $d_1$ the random variable and the observed current data, respectively. When $d_1 = \{Y_C\}$ becomes available, following Bayes' theorem, the corresponding posterior distribution of $\theta_C$ given current and historical data, $g{\scalebox{0.5}{MAP}}(\theta_C|d_1, d_0)$, can be decomposed into two components
\begin{equation*}
	g_{\scalebox{0.5}{MAP}}(\theta_C|d_1, d_0) \propto m(d_1|\theta_C) \cdot f_{\scalebox{0.5}{MAP}}(\theta_C|d_0),
\end{equation*}
where $m(d_1|\theta_C)$ is the binomial likelihood of $Y_C$. For notation simplicity, we hereafter drop the conditional arguments $d_0$ and $d_1$ whenever it does not cause confusion. 

As a dynamic borrowing approach, MAP prior offers some level of robustness as the distribution of $f_{\scalebox{0.5}{MAP}}(\theta_C)$ is often heavy-tailed \cite{NeuenschwanderEtal_2010}. Schmidli and colleagues \cite{SchmidliEtal_2014} proposed the robust MAP prior that can improve its operating characteristics in the presence of prior-data conflict. The idea is to add a vague prior component to $f_{\scalebox{0.5}{MAP}}(\theta_C)$ to hedge the situation in which considerable heterogeneity is observed between current and historical data:
\begin{equation*}
	(1 - w_V) \cdot f_{\scalebox{0.5}{MAP}}(\theta_C) + w_V \cdot f_V(\theta_C),
\end{equation*}
where $f_V(\theta_C)$ is a vague prior that leads to little or no borrowing and $w_V \in [0, 1]$ is a pre-specified weight for the vague prior. 

The idea of robustification is conceptually straightforward. The challenge, however, is that the MAP prior $f_{\scalebox{0.5}{MAP}}(\theta_C)$ is usually from an unknown distribution, which means it is difficult to directly sample from $f_{\scalebox{0.5}{MAP}}(\theta_C)$ in Bayesian computation. With an MCMC sample from $f_{\scalebox{0.5}{MAP}}(\theta_C)$, the authors chose to replace $f_{\scalebox{0.5}{MAP}}(\theta_C)$ with its approximation $\widehat{f_{\scalebox{0.5}{MAP}}}(\theta_C)$ that is based on the results by Dalal and Hall\cite{DalalHall_1983}. It was stated that any parametric prior can be satisfactorily approximated by a mixture of conjugate priors. With binomial data $Y_C \sim Bin(n_C,\theta_C)$ where $\theta_C$ is the binomial probability of interest, $f_{\scalebox{0.5}{MAP}}(\theta_C)$ can be approximated by a weighted mixture of $K$ beta distributions, as beta distribution is conjugate to binomial distribution: 
\begin{equation}\label{eqn:MAPhat}
	\widehat{f_{\scalebox{0.5}{MAP}}}(\theta_C) = \sum^K_{k=1} w_k Beta(a_k,b_k),
\end{equation}
where $\sum^K_{k=1} w_k = 1$. The number of components $K$ is recommended not to exceed the number of historical data \cite{LiEtal_2016}, and is usually chosen to be the smallest number that produces an adequate approximation based on a certain criterion, e.g. by Kullback-Leibler divergence or AIC/BIC. Other parameters $w_k, a_k, b_k$ are estimated by EM algorithm. The vague prior $f_V(\theta_C)$ takes the same form with the components in $\widehat{f_{\scalebox{0.5}{MAP}}}(\theta_C)$. For example, with binary data, $f_V(\theta_C)$ can be the standard uniform prior $Beta(1,1)$ or Jefferys prior $Beta(0.5,0.5)$. Formally, we have
\begin{equation}\label{eqn:robust_MAPhat}
	f_{\scalebox{0.5}{rMAP}}(\theta_C) = (1 - w_V) \cdot \widehat{f_{\scalebox{0.5}{MAP}}}(\theta_C)  + w_V \cdot f_V(\theta_C).
\end{equation}
With rMAP prior \eqref{eqn:robust_MAPhat}, the posterior distribution $g_{\scalebox{0.5}{rMAP}}(\theta_C)$ is also in the form of a weighted mixture of conjugate distributions, of which both the weights and component-wise parameters are updated analytically. 

The difficulty to sample from the unknown distribution of $f_{\scalebox{0.5}{rMAP}}(\theta_C)$ can be circumvented by working with the mixture of conjugate priors \eqref{eqn:robust_MAPhat}. Yet another practical challenge in implementing the robust MAP prior is to properly specify the mixture weight $w_V$. The general principle is to set $w_V$ close to 0 if the historical data is believed to be ``similar with'' current data, or vice versa. However, ascertaining the congruence between historical and current data is far from easy at the designing stage. In most cases, multiple values of $w_V$ are experimented in simulations. The amount of scenarios grows considerably when interim analyses are involved as various $w_V$'s might be considered at each interim. Therefore, methods that objectively adjust the extent of borrowing based on observed current data is desirable to boost the acceptance for the notion of historical data borrowing. 

For the binomial endpoint, Li and colleagues \cite{LiEtal_2016} proposed an empirical Bayes MAP (EB-MAP) prior based on the mixture \eqref{eqn:MAPhat} and introduced an additional parameter $\tau \in (0, \infty)$: 
\begin{equation*}
	f_{\scalebox{0.5}{EB-MAP}}(\theta_C) = \sum^K_{k=1} w_k Beta(\frac{a_k}{\tau},\frac{b_k}{\tau}).
\end{equation*}
The additional parameter $\tau$ ensures that the mean of each Beta component remains unchanged, but its variance is altered: when $\tau < 1$, the variance becomes smaller which results in more borrowing, or vice versa. The value of $\tau$ is determined by maximizing the marginal likelihood of current data $Y_C$ once it becomes available. In particular, the marginal likelihood is
\begin{equation*}
    m(Y_C|\tau) = \sum^K_{k=1} w_k \cdot BetaBin(Y_C|n_C, \frac{a_k}{\tau},\frac{b_k}{\tau}),
\end{equation*}
where $BetaBin(\cdot)$ is the probability mass function of Beta-Binomial($n_C, \frac{a_k}{\tau}, \frac{b_k}{\tau}$) distribution. Convergence issues may arise as $\tau$ approaches either 0 or $\infty$ and the authors proposed mitigating procedures to stabilize the numerical optimization. 

\subsection{A Novel Empirical Bayes Robust MAP Prior}
We hereby propose a novel empirical Bayes robust MAP prior framework that allows borrowing historical data in an adaptive manner. The foundation of our empirical Bayes method is the prior predictive p value ($ppp$). Box \cite{Box_1980} discussed how the prior predictive distribution can be used to gauge the compatibility of the data and prior information. In recent research, the $ppp$ has been used for evaluating prior-data conflict and for choosing the optimal discounting parameter\citep{GravestockHeld_2017, NikolakopoulosEtal_2018, BennettEtal_2021}. Let $T(\cdot)$ be a statistic of $D_1$, the two-sided prior predictive p-value in its most general form \cite{NikolakopoulosEtal_2018} is defined as
\begin{equation*}
    ppp(d_1) = 2 \times \min \{ Pr_{D_1|d_0}( T(D_1) \geq T(d_1), Pr_{D_1|d_0}( T(D_1) \leq T(d_1) \},
\end{equation*}
where $D_1|d_0$ is the predictive distribution of $D_1$ given historical data $d_0$. The corresponding predictive distribution of $T(D_1)$ is 
\begin{equation*}
m(T(D_1)|d_0) = \int m(T(D_1)|\theta_C) f_{\scalebox{0.5}{rMAP}}(\theta_C) d\theta_C.    
\end{equation*}
This implies that the $ppp$ is a function of both $d_1$ and the mixture weight $w_V$ in the rMAP prior. Specifically, with binomial data, we have $d_1 = \{ Y_C \}$ and $T(D_1) = D_1$. The predictive prior p-value is then
\begin{equation}\label{eqn:ppp}
    ppp(Y_C, w_V) = 2 \times \min \{ Pr_{D_1|d_0}( D_1 \geq Y_C), Pr_{D_1|d_0}( D_1 \leq Y_C) \}.
\end{equation}

The optimal mixture weight $w_{EB}$ in EB-rMAP prior can be defined as the smallest value among all $w_V$'s that lead to a reasonably large $ppp$. Since a small $ppp$ indicates strong prior-data conflict, we recommend setting the threshold to a large value in principal. This reflects the fact that borrowing is considered only if there is weak evidence for prior-data conflict. We denote this threshold $\gamma \in (0,1)$ and it is the only tuning parameter in EB-rMAP prior. The sample size of current data $n_C$ may also need to be taken into consideration. More discussions on this will be presented in the next section. On the other hand, if $ppp$ never exceeds $\gamma$, $w_{EB}$ is then set to 1 which implies that the vague prior $f_V(\theta_C)$ is used. This happens when observed current data is very different from the historical data. Specifically, we have
\begin{singlespace}
\begin{equation}\label{eqn:w_eb}
    w_{EB} = 
    \begin{cases}
        \min_{w_V} \{ w_V: ppp(Y_C, w_V) \geq \gamma \}, & \text{if $\exists w_V$ s.t. $ppp(Y_C, w_V) \geq \gamma$ }; \\
        1, & \text{if otherwise}.
    \end{cases}
\end{equation}
\end{singlespace}
{\color{black}
Once $w_{EB}$ is determined, the EB-rMAP prior is in the form of a weighted mixture of conjugate priors. Like the methods reviewed in Section 2.1, the posterior distribution corresponding to EB-rMAP prior is also a weighted mixture of conjugate components. Point estimate and inferences can be drawn from the posterior distribution. 
}

The crucial step in EB-rMAP approach is the calculation of $ppp(Y_C, w_V)$, which requires ascertaining the prior predictive distribution of current data $Y_c$ corresponding to the robust MAP prior \eqref{eqn:robust_MAPhat}. The density of the prior predictive distribution for $n_C$ observations is
\begin{equation*}
    m(Y_C|d_0) = \int m(Y_C|\theta_C) \cdot f_{\scalebox{0.5}{rMAP}}(\theta_C) d\theta_C.
\end{equation*}
As discussed previously, $f_{\scalebox{0.5}{rMAP}}(\theta_C)$ is in the form of a mixture of Beta distributions. Following Fubini's theorem, $m(Y_C|d_0)$ is a mixture of Beta-Binomial densities, of which each component corresponds to individual Beta prior and the binomial likelihood $m(Y_C|\theta_C)$. This distribution can be calculated in \textit{RBesT} package \cite{Weber_2021_RBesT}. The probability $Pr_{D_1|d_0}( D_1 \leq Y_C)$ is then computed based on the distribution function of the mixture distribution. Lastly, $ppp(Y_C, w_V)$ is obtained according to \eqref{eqn:ppp}. 

{\color{black}
The optimal weight $w_{EB}$ is determined by multiple factors including the distribution of MAP prior, the vague prior, current trial sample size ($n_C$) and the threshold $\gamma$. However, we can mostly focus on setting the threshold $\gamma$ for EB-rMAP prior in practice for the following reasons. First, the historical data are pre-selected, and hence the MAP prior is ``fixed'' barring randomness in MCMC. Second, the choice of vague prior is generally obvious such as $Beta(1,1)$ with binomial data. Finally, it is most likely that there are limited proposals for $n_C$, and we also observe that the behavior of $w_{EB}$ is fairly insensitive to the choice of $n_C$. To help determine $\gamma$, we calculate $w_{EB}$'s with various $\gamma$ at different observed current trial data. With binomial data, for example, it is possible to enumerate all possible values of current data ($Y_C = 0, 1,\dots, n_C$). We then examine $w_{EB}$'s at selected $Y_C$'s to identify the $\gamma$ that yields a desirable level of borrowing. More illustrations on this matter will be provided in Simulation and Data Analysis sections. Simulation studies evaluating design operating characteristics corresponding to different $\gamma$'s may also be needed. 
}

The EB-rMAP method differs from the EB-MAP \cite{LiEtal_2016} in several aspects. First, EB-rMAP is built on robust MAP prior \cite{SchmidliEtal_2014}, while EB-MAP modifies the original MAP prior \cite{NeuenschwanderEtal_2010}. Secondly, the optimization procedures are very different albeit both methods are empirical Bayes in nature. The parameter being optimized in EB-rMAP method is the mixture weight $w_V$ bounded within $[0,1]$. In contrast, there is no upper limit for its counterpart $\tau$ in EB-MAP. Therefore, the optimization in our method could be more computationally tractable such that no additional rule is needed to ensure convergence. One can create a fine grid of $w_V$, evaluate the $ppp(d_1, w_V)$ at each value and determine $w_{EB}$ following \eqref{eqn:w_eb}. Finally, the EB-rMAP method provides a unified framework that applies to most commonly encountered endpoints as listed in Table \ref{tab:endpoints}. In particular, the important time-to-event (TTE) endpoint can be modeled as count data under the constant hazard assumption so that EB-rMAP prior is applicable. We elaborate the EB-rMAP prior for TTE endpoint in the next subsection.
\begin{table}[htb]
    \centering
    \caption{Applicable Scenarios for EB-rMAP Prior}
    \begin{tabular}{cccc}
    Type of Endpoint & Likelihood & Prior & Prior Predictive \\
    \hline
    Binary & Binomial & Beta & Beta-Binomial \\
    Continuous & Normal (known SD) & Normal & Normal \\
    Count & Poisson & Gamma & Poisson-Gamma \\
    Time-to-event$\dagger$ & Poisson & Gamma & Poisson-Gamma \\
    \hline
    \multicolumn{4}{l}{$\dagger$ Under constant hazard assumption}
    \end{tabular}
    \label{tab:endpoints}
\end{table}

\subsection{EB-rMAP Prior with Time-To-Event Endpoint}
{\color{black}
Compared to continuous and binary endpoints, historical data borrowing with TTE endpoint has received less attention until recently \cite{RoychoudhuryNeuenschwander_2020, SmithEtal_2020}. Let $h=1,\cdots,H$ index the historical trials. The historical TTE data are summarized in two quantities: number of events and total at-risk time (exposure), denoted respectively by $r_h$ and $exp_h$. 

The EB-rMAP approach is applicable to TTE data by assuming a constant hazard rate $\lambda_h$ for each trial, or equivalently, an exponential distribution (with rate $\lambda_h$) for TTE. This assumption gives rise to an equivalent Poisson model formulation for the number of events $r_{h}$:
\begin{equation}\label{eqn:TTE_Poisson}
    r_{h} \sim Poisson(\lambda_{h} exp_{h}).
\end{equation}
The expectation of $r_h$ is the Poisson rate $\lambda_{h} exp_{h}$. Taking log transformation, the Poisson model is
\begin{equation*}
    \log (E[r_{h}]) = \log\lambda_{h} + \log (exp_h)
\end{equation*}
where the log-exposure, $\log(exp_h)$, is termed the offset. The original MAP prior assumes that the log-hazard rates arise from a common normal distribution:
\begin{equation}\label{eqn:TTE_MAP}
    \log\lambda_{1},\cdots,\log\lambda_{H} \sim N(\log\lambda, \tau^2).
\end{equation}
The mean log-hazard rate $\log\lambda$ generally has a weakly-informative normal prior, and $\tau^2$ is the exchangeability parameter that regulates the amount of borrowing. 

Given the Poisson likelihood for $r_{h}$'s, the MAP prior for $\lambda$ may be approximated by a weighted mixture of Gamma priors. The vague prior to construct the robust MAP prior could be a Gamma prior with an effective sample size of 1. Note that in the case of TTE endpoint, the effective sample size refers to the number of events, instead of the number of subjects as with binary and normal endpoints. Denote $r_{C}$ and $exp_C$ the number of events and total exposure in current trial, respectively. The EB-rMAP weight $w_{EB}$ can be determined using the same procedure described in the previous subsection with current data. The posterior distribution corresponding to EB-rMAP, from which point estimate and inference are obtained, is also a weighted mixture of Gamma components. The aforementioned tasks can be carried out using \textit{RBesT} package as the Gamma-Poisson conjugate model is implemented.

One challenge that pertains to all historical borrowing methods with a TTE endpoint lies in obtaining the historical data. While the number of events $r_h$ may be available at the study summary level, the total exposure $exp_h$ can only be calculated using individual level data which may not always be available. This challenge can be mitigated by adopting algorithms and/or tools \cite{ParmarEtal_1998, LiuEtal_2021_ipdfromkm} that reconstruct individual level data from published Kaplan-Meier survival curves. Meanwhile, ascertaining the total exposure in current trial $exp_C$ at the design stage is a more difficult task, compared to setting the sample size $n_C$ in cases of binary and normal endpoints. However, we noticed that the behavior of $w_{EB}$ is insensitive to the value of $exp_C$. Therefore, a ballpark value of $exp_C$ can be obtained via a simple simulation that generates individual level data for the current trial, and it will likely suffice the purpose of calibrating EB-rMAP prior at the design stage. 

In certain applications, it might be overly simplistic to assume a constant hazard. A more flexible model is the piecewise exponential (PWE) model. The PWE model partitions the follow-up period into mutually exclusive time intervals, and the hazard rate within each time interval is assumed to be constant. The EB-rMAP prior framework described in this subsection can be applied independently to each time interval. In particular, the threshold $\gamma$ can be different across time intervals, determined by factors such as the total exposure within the time interval. Some practical considerations regarding PWE setting are provided in the Concluding Remarks section. 
}

\section{Simulation Studies}
\subsection{Normal Endpoint with Known Standard Deviation}
Under the single arm setting, we evaluate the performance of EB-rMAP in simulations. The historical data to derive the original MAP prior are from five historical trials in moderate to severe Crohn's disease \cite{HueberEtal_2012}, available in \textit{crohn} dataset of \textit{RBesT} package. The sample sizes ranged from 20 to 328, and the continuous endpoint was the change from baseline in Crohn's Disease Activity Index (CDAI) at week 6. An improvement in outcome corresponds to a negative change from baseline. We assume the common standard deviation of the endpoint $\sigma = 40$. A random effect meta-analysis yields a point estimate of -46.8 for the mean response, \textcolor{black}{with 95\% confidence interval (-55.1, -38.6)}. The historical means $\theta_1,\cdots,\theta_5$ are assumed to follow a normal distribution $N(\mu_C, \sigma_C^2)$. To derive the original MAP prior, we posit a weakly-informative prior $N(-50,40^2)$ for $\mu_C$ as data would be sufficiently informative for it. The exchangeability parameter $\sigma_C$ has a half-normal prior $HN(s=5)$. This configuration results in a fairly informative MAP prior with an effective sample size (ESS, by the definition of Neuenschwander et al. \cite{NeuenschwanderEtal_2020_ESS}) of 31.9. The vague component $f_V(\mu_C)$ to construct the robust MAP prior is $N(-50, 40^2)$ which has an ESS of 1. 

\begin{figure}
	\centering
	\includegraphics[width=\textwidth, keepaspectratio]{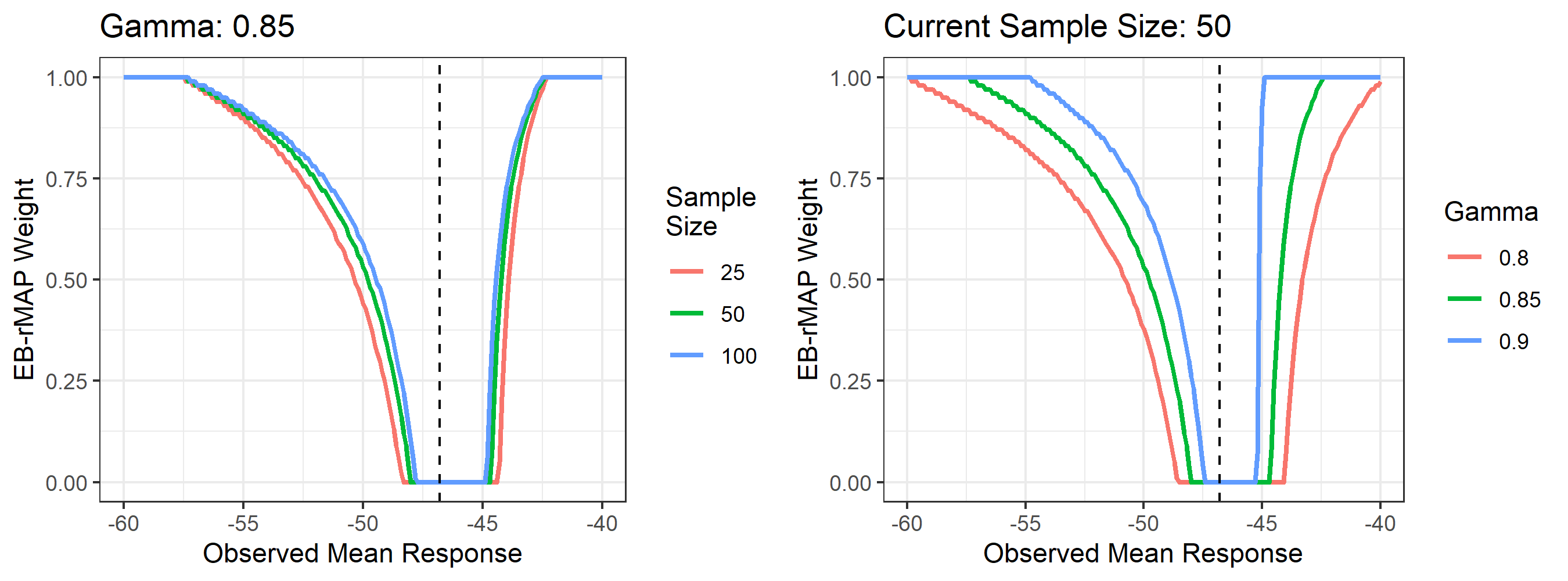}
	\caption{EB-rMAP Weights versus Observed Mean Responses in Current Data}
	\label{fig:EBweights}
\end{figure}

We first investigate the behavior of EB-rMAP weight $w_{EB}$ and how it is impacted by the current data sample size $n_C$ and the tuning parameter $\gamma$. Figure \ref{fig:EBweights} shows the values of $w_{EB}$ against different values of observed mean response in current data. In the left panel, the curves correspond to different current sample sizes $n_C = 25/50/100$ respectively, while fixing all other specifications. Similarly, in the right panel, we only vary $\gamma$'s (0.8, 0.85 and 0.9). The meta-analysis point estimate of historical data is marked by the vertical dashed line. There is generally a window around the historical mean response within which the EB-rMAP weight is lower than 1. The weight tends to 0 which indicates the informative MAP prior is essentially used, when the observed current and historical mean responses are close. The method puts more weights on the vague prior as the two mean responses gets further apart. The further the prior-data conflict, the closer $w_{EB}$ approaches 1. These are very favorable behaviors because $w_{EB}$ responds properly to prior-data conflict, or the lack thereof. Fixing all other aspects, the window is wider with a smaller $\gamma$ due to a less stringent requirement of agreement. On the other hand, although the window tends to be wider with a smaller $n_C$, its impact is fairly limited in this configuration. 
{\color{black}
In the right panel of Figure \ref{fig:EBweights}, with $\gamma = 0.8$, $w_{EB}$ tends to 1 when current mean response is below -60, which is outside the 95\% CI of historical meta-analysis. Therefore, it may be considered that borrowing with $\gamma = 0.8$ is too aggressive. The same arguments apply when $\gamma = 0.85$, although it is slightly more conservative. Therefore, we set the threshold $\gamma$ to 0.9 for EB-rMAP prior in this simulation. 
}

With the original MAP and vague priors, we design a current single arm study with sample size $n_C = 50$. We vary the true current mean response between -55 and -35. With each simulated current mean response $d_1=\Bar{y}_C$, the study is deemed successful if the decision rule is met:
\begin{equation*}
Pr(\mu_C < -40|d_0, d_1) > 0.95.    
\end{equation*}
We compare our EB-rMAP approach with robust MAP priors with fixed $w_V = 0, 0.5, 1$ respectively. Recall that the scenario of $w_V = 0$ is essentially the original MAP prior, while the vague prior is used with $w_V = 1$. The operating characteristics for evaluation are probability-of-success (PoS)\cite{Chuang_2006}, absolute bias and mean square error of the posterior median estimator for $\mu_C$. 

\begin{figure}
	\centering
	\includegraphics[width=\textwidth, keepaspectratio]{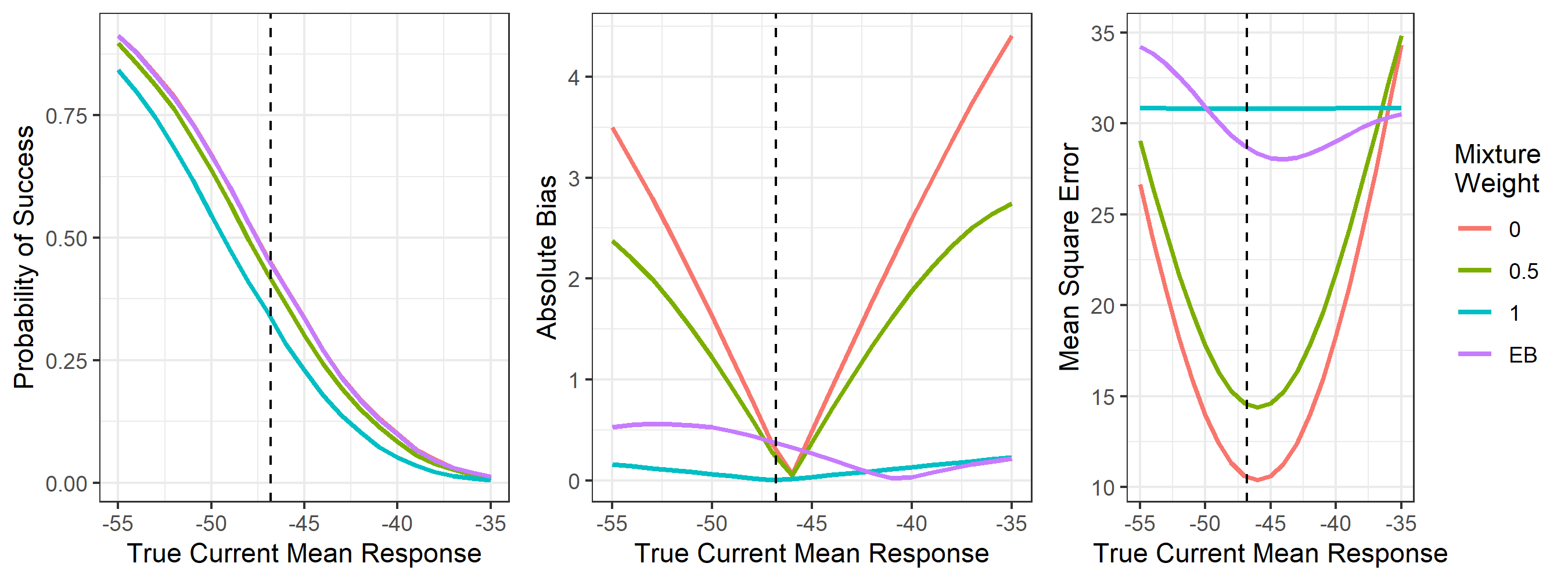}
	\caption{Simulation: Normal Endpoint with Known Standard Deviation}
	\label{fig:sim_normal}
\end{figure}

Simulation results are presented in Figure \ref{fig:sim_normal} based on 5,000 iterations at each value of true current mean response. The PoS is well-maintained with EB-rMAP as the PoS curve almost overlaps with that of the original MAP ($w_V$ = 0). On the other hand, the absolute bias of EB-rMAP is similar to that of the vague prior ($w_V$ = 1) which is expected to yield the smallest bias. EB-rMAP prior is also comparable with vague prior in terms of MSE. 

\subsection{Binary Endpoint}
In the simulation with binary endpoint, we use \textit{AS} dataset from \textit{RBesT} package which contains data from 8 clinical trials in ankylosing spondylitis. Improvement in ankylosing spondylitis is assessed by Assessment of SpondyloArthritis International Society (ASAS) score containing four domains. The binary efficacy endpoint is ASAS20 at week 6, which is defined as an improvement of at least 20\% and an absolute improvement of at least 1 unit (on a 0-10 scale) in at least three of four domains, with no worsening of the remaining domain. A meta-analysis yields a point estimate of 0.25 (95\% CI: 0.20, 0.31) for the ASAS20 rates from historical data. 
	
Denote $p$ the ASAS20 rate. The original MAP prior method assumes that the log-odds of ASAS20 rates in historical and current studies arise from a common normal distribution $N(\mu_C, \sigma_C^2)$, where $\mu = \log(p/(1-p))$. We posit a $HN(0.5)$ prior for $\sigma_C$, and the overall mean logit $\mu_C$ has a normal prior $N(0,2^2)$ which is considered weakly-informative on the log-odds scale. The derived MAP prior has an ESS of 37.7. Since the original MAP prior for a binary endpoint is approximated by a mixture of Beta distributions, the vague component to construct robust MAP prior should also be a Beta prior and we use $Beta(1,1)$ (ESS = 2) in the simulation. 

We compare EB-rMAP approach ($\gamma = 0.8$) with robust MAP prior with fixed weight $w_V = 0, 0.5, 1$, using the same three operating characteristics as the previous subsection. The current single arm study has a sample size $n_C$ of 50. We perform 5,000 iterations at each true current ASAS20 rate considered, ranging from 0.20 to 0.32. The decision rule to claim trial success is $Pr(p_C > 0.2 |d_0, d_1) > 0.9$. 

\begin{figure}
	\centering
	\includegraphics[width=\textwidth, keepaspectratio]{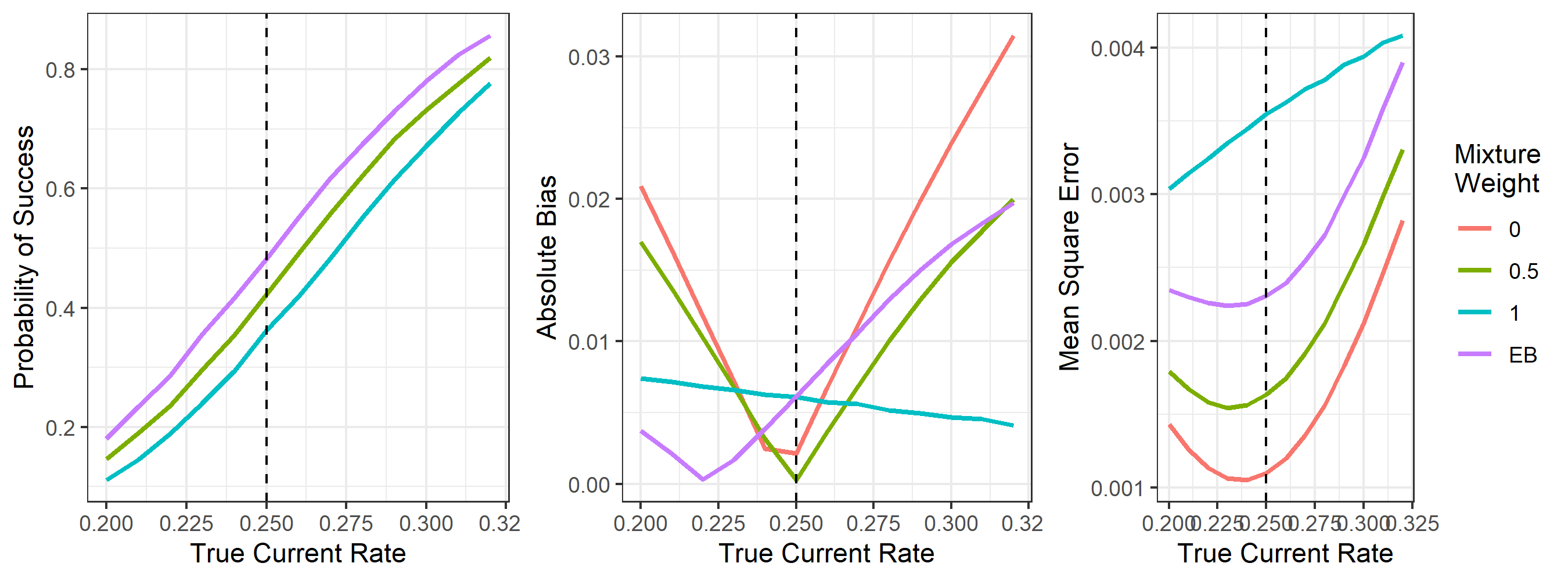}
	\caption{Simulation: Binary Endpoint}
	\label{fig:sim_binary}
\end{figure}

Results presented in Figure \ref{fig:sim_binary} suggest that EB-rMAP approach strikes a good balance between PoS and estimation quality. It has nearly identical PoS with the original MAP prior. Meanwhile, its bias and MSE are mostly between those of original MAP prior and the vague prior. 

\subsection{Time-to-Event Endpoint}
In this subsection, we evaluate the performance of EB-rMAP prior with TTE endpoint using simulated historical data. Without loss of generality, we set $K=1$ in PWE model. This essentially assumes the TTE is exponentially distributed so that the relationship between hazard rate $\lambda$ and median overall survival (mOS) is: $\lambda = \log(2)/\mbox{mOS}$. We consider four historical studies and fix their total exposures (in years) at 5, 10, 15 and 20, respectively. The numbers of deaths in each trial are simulated from a Poisson distribution according to \eqref{eqn:TTE_Poisson}, where the hazard rate is set to 0.4 for all four historical studies. This translates to a median overall survival of 1.73 years. For the current trial, the total exposure is fixed at 30 years, and we consider different hazard rates ranging from 0.3 to 0.7 (mOS from 1 to 2.31 years). At each current hazard rate, 1,000 simulations are conducted. 

The hyper-priors in MAP prior \eqref{eqn:TTE_MAP} are
\begin{equation*}
    \log\lambda \sim N(0, 10^2); \tau \sim HN(0.5).
\end{equation*}
The normal prior is very weakly-informative for log-hazard rate $\log\lambda$. The $HN(0.5)$ prior covers small to large between-trial variability and therefore corresponds to moderate borrowing. The mean of the vague Gamma prior in robust MAP prior, whose effective number of events is 1, is determined by the median of the original MAP prior. We compare EB-rMAP approach ($\gamma = 0.75$) to robust MAP with $w_V = 0, 0.5,1$ in the same three metrics as before: PoS, absolute bias and MSE. The PoS is corresponding to the decision rule $Pr(\lambda \leq 0.5|d_0, d_1) > 0.9$. 

Simulation results are reported in Table \ref{tab:sim_TTE} for the five current hazard rates considered. When the current hazard rate $\lambda_C \leq 0.5$, it can be seen that the operating characteristics of EB-rMAP prior are generally between those of the original MAP ($w_V = 0$) and vague prior ($w_V = 1$). On the other hand, when $\lambda_C > 0.5$, meeting the decision rule leads to an erroneous conclusion. In this case, the PoS (error rate) and bias of EB-rMAP are comparable to those of the vague prior. This showcases the robustness of the EB-rMAP prior. 
\begin{table}[htbp]
  \centering
  \caption{Simulation: TTE Endpoint}
    \begin{tabular}{c|ccc|ccc|ccc}
          & \multicolumn{3}{c}{$\lambda_C=$0.3} & \multicolumn{3}{c}{$\lambda_C=$0.4} & \multicolumn{3}{c}{$\lambda_C=$0.5} \\
    \hline
    $w_V$     & PoS   & Abs. Bias$\dagger$ & MSE$\dagger$   & PoS   & Abs. Bias & MSE   & PoS   & Abs. Bias & MSE \\
    \hline
    EB    & 0.705 & 18.8  & 9.1   & 0.346 & 7.3  & 10.6  & 0.129  & 3.7  & 14.3 \\
    0     & 0.738 & 27.4  & 6.5   & 0.440  & 8.1   & 6.9   & 0.189 & 42.9  & 10.9 \\
    0.5   & 0.742 & 13.3  & 7.6   & 0.416 & 9.4   & 8.4   & 0.171 & 33.4  & 12.3 \\
    1     & 0.704 & 8.1   & 9.9   & 0.332 & 9.2   & 12.9  & 0.112 & 11.3  & 16.4 \\
    \hline
          & \multicolumn{3}{c}{$\lambda_C=$0.6} & \multicolumn{3}{c}{$\lambda_C=$0.7} &       &       &  \\
    \hline
      $w_V$ & PoS   & Abs. Bias & MSE   & PoS   & Abs. Bias & MSE   &       &       &  \\
    \hline
    EB    & 0.048 & 12.4  & 18.6  & 0.009 & 15.6  & 23.2  &       &       &  \\
    0     & 0.073 & 72.7  & 17.8  & 0.026 & 91.8  & 26.7  &       &       &  \\
    0.5   & 0.061 & 52.5  & 18.5  & 0.015 & 60.8  & 26.0    &       &       &  \\
    1     & 0.038 & 15    & 19.7  & 0.006 & 16.4  & 23.7  &       &       &  \\
    \hline
    \multicolumn{10}{l}{$\dagger$ Absolute bias and MSE are multiplied by 1000.}
    \end{tabular}%
  \label{tab:sim_TTE}%
\end{table}%

\section{Data Analysis with Time-To-Event Data}
{\color{black}
In this section, we illustrate the EB-rMAP approach for the TTE endpoint via re-analyzing the data from 10 oncology studies used in Roychoudhury and Neuenschwander \cite{RoychoudhuryNeuenschwander_2020}. The full data are presented in Appendix A. Nine studies are regarded as the historical data, while the remaining one is the current data. Originally, the four year follow-up period was partitioned into 12 intervals. In our analysis to illustrate EB-rMAP prior, we combine the data across time intervals as shown in Table \ref{tab:analysis_data}. 

\begin{table}[htbp]
  \centering
  \caption{Historical and Current Data in Section 4 \\ (Number of Events/Total Exposure in Years)}
    \begin{tabular}{|c|c|c|c|c|}
    \hline
    Historical 1 & Historical 2 & Historical 3 & Historical 4 & Historical 5 \\
    \hline
    14/45 & 32/110.8 & 29/114.7 & 13/25.3 & 22/23.7 \\
    \hline
    Historical 6 & Historical 7 & Historical 8 & Historical 9 & Current \\
    \hline
    31/86.4 & 18/36.7 & 10/48.7 & 10/25.4 & 32/117.6 \\
    \hline
    \end{tabular}%
  \label{tab:analysis_data}%
\end{table}%

The original MAP prior \eqref{eqn:TTE_MAP} for $\lambda$ is derived using the nine historical trials: 
\begin{equation*}
    \log\lambda_{1},\cdots,\log\lambda_{9} \sim N(\log\lambda, \tau^2),
\end{equation*}
where
\begin{equation*}
    \log\lambda \sim N(0, 10^2); \tau \sim HN(0.5).
\end{equation*}
The hyper-prior for $\tau$ enables moderate borrowing, and the effective number of events is 15.3. The MAP prior is approximated by two gamma components (mean and number of observations, or ``mn'' parameterization): 
\begin{equation*}
    \widehat{f_{\scalebox{0.5}{MAP}}}(\lambda) = 0.82*Ga(0.37, 21.4) + 0.18*Ga(0.62, 3.8).
\end{equation*}

The vague gamma prior $f_V(\lambda)$ to construct robust MAP prior is $Ga(0.42, 1)$. It has an effective number of events of 1 and its mean equals to that of the MAP prior. The left panel in Figure \ref{fig:analysis1} shows the densities of MAP and the vague prior respectively. The vague prior clearly covers a wider range of possible values of hazard rate, and hence should be more robust. 

The total exposure was 117.6 years in the current trial. We use the right panel in Figure \ref{fig:analysis1} to help determine $\gamma$ for EB-rMAP prior, in which values of $w_{EB}$ corresponding to different number of events observed in current trial ($r_C$) are shown. The vertical dashed line indicates $r_C$ that yields a hazard rate of 0.37, the mean hazard rate from historical meta-analysis. In the current trial, there were 32 events and the $w_{EB}$ is 0.47, 0.54 and 0.62 for $\gamma = 0.85, 0.9$ and 0.95, respectively. We hereafter use $\gamma = 0.9$ in the analysis. 
\begin{figure}
	\centering
	\includegraphics[width=\textwidth, keepaspectratio]{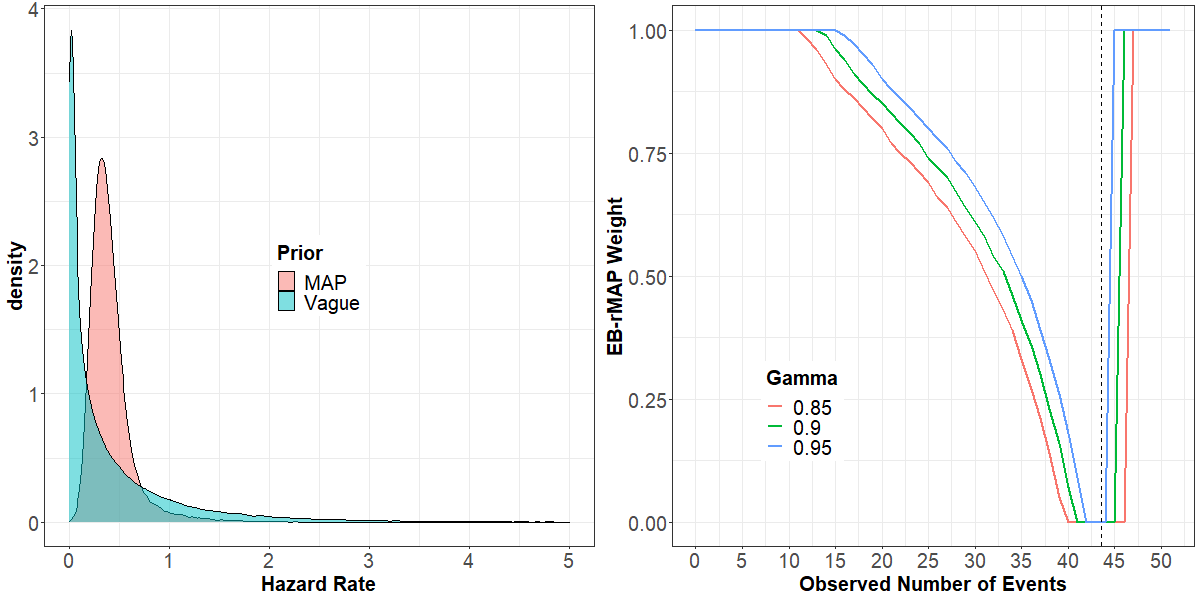}
	\caption{Left: Densities of Prior Distributions; \\ Right: $w_{EB}$ Corresponding to Different $\gamma$ (117.6 Years Total Exposure)}
	\label{fig:analysis1}
\end{figure}

In addition to EB-rMAP prior, two other models were implemented: models with the original MAP prior and the vague $Ga(0.42, 1)$ prior. The posterior median hazard rates and 95\% credible intervals are reported in Table \ref{tab:analysis}. For EB-rMAP prior, its posterior median hazard rate and width of credible interval are in-between the counterparts in the other two models. Finally, Figure \ref{fig:haz_density} shows the posterior distributions of hazard rate by each method. 

\begin{table}[htbp]
  \centering
  \caption{Posterior Median and 95\% Credible Interval}
    \begin{tabular}{|c|c|c|}
    \hline
    EB-rMAP & MAP   & Vauge \\
    \hline
    0.281 (0.199, 0.384) & 0.285 (0.203, 0.386) & 0.270 (0.187, 0.375) \\
    \hline
    \end{tabular}%
  \label{tab:analysis}%
\end{table}%

\begin{figure}
	\centering
	\includegraphics[keepaspectratio]{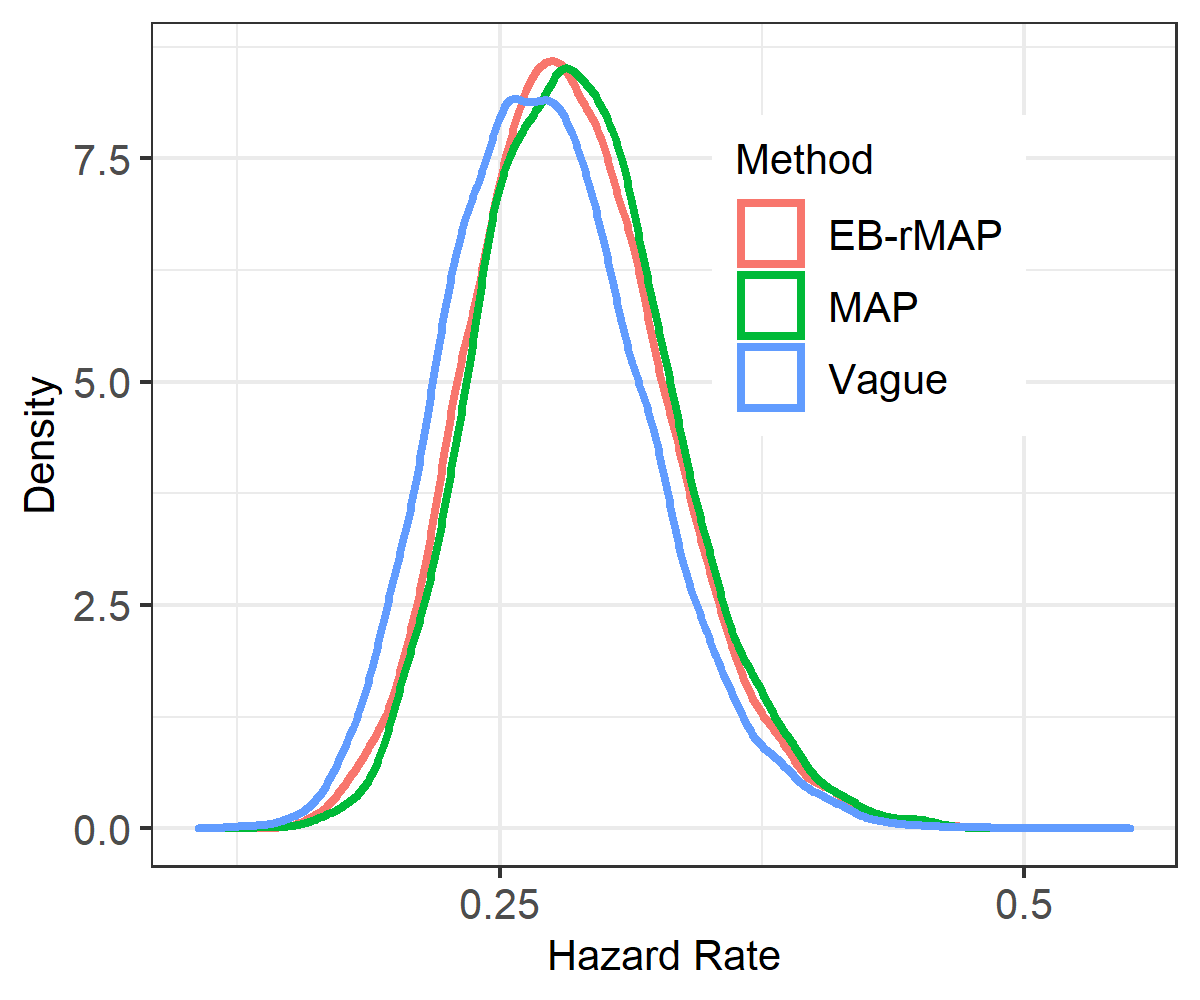}
	\caption{Posterior Distributions of Hazard Rate in Interval (1.25, 1.50] Years}
	\label{fig:haz_density}
\end{figure}
}

\section{Concluding Remarks}
{\color{black}
Historical data of previous studies with similar enrollment criteria are often available in the analysis of current clinical trial.  Several Bayesian methods have been proposed for including historical data as prior information, such as the power prior, the commensurate power prior, the (robust) MAP prior and the empirical Bayes MAP prior. 
}
We propose a novel empirical Bayes robust MAP prior to address an important practical challenge in implementing robust MAP prior: to properly specify the mixture coefficient $w_V$. Built upon the Box's prior predictive p-value that quantifies prior-data conflict, our EB-rMAP framework allows adaptively borrowing historical data in a data-dependent manner. The computation in EB-rMAP prior utilizes existing software packages which greatly reduces the amount of programming. R programs are avaliable on GitHub: (link will be provided upon manuscript acceptance). The optimization procedure can be more stable and efficient than existing empirical Bayes methods as the empirical Bayes mixture coefficient $w_{EB}$ is bounded within $(0, 1)$. Simulation studies suggest that the EB-rMAP method strikes a good balance between maintaining probability-of-success and yielding robust point estimates. Our unified framework seamlessly applies to most commonly encountered data types. Compared to the empirical Bayes MAP prior \cite{LiEtal_2016}, the novel EB-rMAP prior has one additional tuning parameter $\gamma$ that need to be prespecified. While this grants more flexibility, extensive simulation studies are required to choose the value for $\gamma$ that yields satisfactory operating characteristics. 

{\color{black}
When a PWE model is assumed for TTE data, the EB-rMAP prior procedure can be applied to each time interval. Since historical and current data vary across intervals, it might be necessary that the threshold $\gamma$'s are interval-specific. This approach ignores the between-interval correlation, and therefore may be considered sub-optimal. Roychoudhury and Neuenschwander \cite{RoychoudhuryNeuenschwander_2020} proposed a flexible model that accounts for such correlation. Their model cannot be implemented in \textit{RBesT} package and is more computational intensive. Additionally, there may be special considerations pertaining not only to EB-rMAP prior, but all methods borrowing TTE data. One example is how the intervals should be determined when follow-up times differ among studies, as the results could be sensitive to the partition. 
}

The endpoints considered in our simulation studies and data analysis are all efficacy-oriented. Meanwhile, an important binary safety endpoint often encountered in phase 1 dose escalation studies is the presence of dose-limiting toxicity (DLT). Bayesian logistic regression model \cite{NeuenschwanderEtal_2008BLRM} (BLRM) is a popular model-based approach to fit cumulative DLT data and to make dose recommendation for the next cohort of patients. The model has two parameters, namely intercept and slope, for which a bivariate normal prior is posed. Historical information on the dose-toxicity profile may be incorporated via an MAP prior \cite{NeuenschwanderEtal_2015}. The operating characteristics of BLRM can depend heavily on the bivariate normal prior due to the limited sample size in dose escalation trials. Therefore, properly handling of prior-data conflict is crucial to ensure the robustness. The idea of robust MAP prior has been extended to BLRM via the exchangeable-non-exchangeable (EX-NEX) model \cite{NeuenschwanderEtal_2016}. Pre-specifying the weights for the EX or NEX model is critical and follows the same principle with the rMAP prior. However, it is challenging to apply the EB-rMAP prior method in this setting because it is not obvious how to obtain the mixture that approximates the original MAP prior under BLRM. Future research is needed to explore alternative approaches to quantify the prior-data conflict. 

The EB-rMAP prior adjusts the extent of borrowing based on the agreement between historical and current trials. As the real world evidence (RWE) becomes increasingly important in recent years, information from external sources, especially observational studies, is brought into the historical data borrowing paradigm. In addition to the discrepancy in responses, the distributions in baseline covariates from external sources may be different from those in the current trial. Ignoring such imbalances may lead to bias and erroneous conclusions. This fact prompts new methods being proposed to take the baseline covariates into consideration when adjusting the extent of borrowing \cite{WangEtal_2019, LiuEtal_2021_PSMAP, LinEtal_2021}. It is of future research interest to explore whether EB-rMAP prior can be implemented in combination with such methods to yield more robust analyses with RWE. 

{\color{black}
Although empirical Bayes MAP and EB-rMAP prior approaches have proven to be very effective, they are by no means a panacea, and continued reach and development is needed. Indeed there are needs in developing hierarchical analyses that are efficient and effective, but also robust with respect to prior information and other characteristics. If such robustness is not present, or the prior-data conflict is not properly addressed, then the sources of the sensitivity must be investigated and questioned.  Broadening our knowledge of statistical influence of the hyper-prior and deepening our understanding of the limitations are important aspects of these developments. 
}

\subsection*{Data Availability Statement}
Data sharing not applicable to this article as no new datasets were generated or analyzed during the current study. 

\bibliography{references}


\appendix
\section{Data Used in the TTE Analysis Section}
Roychoudhury and Neuenschwander \cite{RoychoudhuryNeuenschwander_2020} extracted the data from 10 published Kaplan-Meier curves using the Parmar et al. technique \cite{ParmarEtal_1998}. The follow-up period was divided into 12 time intervals (in years). The first nine studies were regarded as historical trials, while the tenth one current trial. The number of events and total exposure in years are reported in the table below. 

\begin{table*}[h]
    \centering
    \caption{Data from Roychoudhury and Neuenschwander \cite{RoychoudhuryNeuenschwander_2020}}
\begin{tabular}[t]{|c|c|c|c|c|c|c|c|c|c|c|}
\hline
Interval & 1 & 2 & 3 & 4 & 5 & 6 & 7 & 8 & 9 & Current\\
\hline
0.00-0.25 & 1/9.4 & 9/21.1 & 1/21.9 & 1/5.6 & 5/6.4 & 0/17.8 & 2/8 & 0/9.2 & 2/5.2 & 1/23.4\\
\hline
0.25-0.50 & 3/8.8 & 1/19.9 & 3/21.4 & 2/5.2 & 3/5.4 & 6/17 & 2/7.5 & 1/9.1 & 0/5 & 5/22.6\\
\hline
0.50-0.75 & 3/7.9 & 0/19.8 & 5/20.4 & 2/4.8 & 6/4.2 & 3/15.9 & 5/6.6 & 3/8.6 & 3/4.6 & 17/19.9\\
\hline
0.75-1.00 & 4/7 & 10/18.5 & 7/18.9 & 4/4 & 2/3.2 & 12/14 & 3/5.6 & 4/7.8 & 1/4.1 & 0/17.8\\
\hline
1.00-1.25 & 3/6.1 & 6/16.5 & 9/16.9 & 3/3.1 & 3/2.6 & 8/11.5 & 3/4.9 & 1/7.1 & 4/3.5 & 2/17.5\\
\hline
1.25-1.50 & 0/5.8 & 6/15 & 4/15.2 & 1/2.6 & 3/1.9 & 2/10.2 & 3/4.1 & 1/6.9 & 0/3 & 7/16.4\\
\hline
1.50-1.75 & 0/5.8 & 5/13.6 & 5/14.1 & 3/2.1 & 0/1.5 & 3/9.6 & 2/3.5 & 4/6.2 & 1/2.9 & 8/14.5\\
\hline
1.75-2.08 & 2/7.3 & 9/15.7 & 10/16.2 & 0/2.3 & 2/1.7 & 2/11.9 & 3/3.8 & 1/7.4 & 1/3.5 & 4/17.2\\
\hline
2.08-2.50 & 0/8.8 & 9/16.2 & 0/18.5 & 0/2.9 & 1/1.5 & 11/12.4 & 3/3.6 & 6/8 & 0/4.2 & 0/21\\
\hline
2.50-2.92 & 6/7.6 & 3/13.6 & 0/18.3 & 0/2.9 & 1/1 & 1/9.9 & 0/2.9 & 0/6.7 & 0/4.2 & 6/19.7\\
\hline
2.92-3.33 & 0/6.2 & 0/12.5 & 3/17 & 0/2.9 & 1/0.6 & 0/9.4 & 0/2.9 & 0/6.6 & 0/4.1 & 2/17.4\\
\hline
3.33-4.00 & 0/10 & 0/20.1 & 7/24.5 & 0/4.7 & 0/0.7 & 10/12.1 & 0/4.7 & 0/10.7 & 0/6.7 & 0/27.5\\
\hline
\end{tabular}
\end{table*}
\end{document}